# Artificial Bone and Teeth through Controlled Ice Growth in Colloidal Suspensions

Antoni P. Tomsia, Eduardo Saiz, and Sylvain Deville

# Artificial Bone and Teeth through Controlled Ice Growth in Colloidal Suspensions

Antoni P. Tomsia[1], Eduardo Saiz[1] and Sylvain Deville[2]

[1] *Materials Sciences Division, Lawrence Berkeley National Laboratory, One Cyclotron Road, Berkeley, CA 94720, USA*
[2] *Laboratory of Synthesis and Functionalisation of Ceramics, FRE2770/Saint-Gobain CREE, 550 Ave. Alphonse Jauffret, 84306 Cavaillon Cedex, France*

**Abstract.** The formation of regular patterns is a common feature of many solidification processes involving cast materials. We describe here how regular patterns can be obtained in porous alumina and hydroxyapatite (HAP) by controlling the freezing of ceramic slurries followed by subsequent ice sublimation and sintering, leading to multilayered porous ceramic structures with homogeneous and well-defined architecture. These porous materials can be infiltrated with a second phase of choice to yield biomimetic nacre-like composites with improved mechanical properties, which could be used for artificial bone and teeth applications. Proper control of the solidification patterns provides powerful means of control over the final functional properties. We discuss the relationships between the experimental results, ice growth fundamentals, the physics of ice and the interaction between inert particles and the solidification front during directional freezing.

## INTRODUCTION

Materials that are strong, ultra-light weight and tough are in demand for a range of applications and biomaterials, such as for bone or tooth replacement, can be included in this category [1-3]. The ideal bone substitute is not a material that interacts as little as possible with the surrounding tissues, but one that will form a secure bond with the tissues by allowing, and even encouraging new cells to grow and penetrate. One way to achieve this is to use a material that is osteophilic and porous, so that new tissue, and ultimately new bone, can be induced to grow into the pores and help prevent loosening and movement of the implant. Resorbable bone replacements have been developed from inorganic materials that are very similar to the apatite composition of natural bone.

In recent years, considerable attention has been given to the development of fabrication methods to prepare porous ceramic scaffolds for osseous tissue regeneration [4-7]. The ideal fabrication technique should produce complex-shaped scaffolds with controlled pore size, shape and orientation in a reliable and economical way. However, all porous materials have a common limitation: the inherent lack of strength associated with porosity. Hence, their application tends to be limited to low-stress locations, such as broken jaws or fractured skulls. Therefore, the unresolved dilemma is how to design and create a scaffold that is both porous and strong, and how to derive strong and tough composites from these scaffolds. This could be achieved by designing layered materials. While the potential of layered materials has long been recognized [8,9], their creation requires solving a two-fold problem, namely the design of optimum microstructures, and development of fabrication procedures to implement these designs. Inspiration from natural materials (biomimetics) has recently appeared as a promising source of stimulus to solve these issues.

## Nacre as a Blueprint

Nature is far more successful in designing strong and tough lightweight materials than is found with conventional engineering composites. Bone and nacre, for example, are sophisticated composites whose unique combination of mechanical properties derives from an architectural design that spans from the nanometer scale to the millimetre, or even centimetre, scale [10-13]. Unlike engineering composites where the properties are invariably governed by the "rule of mixtures", i.e., the properties of the composite are intermediate between those of its constituents, in many natural materials the properties of the composite are often many times, even orders of magnitude, greater than either of its constituent phases. Few structural engineering materials have such a hierarchy of structure, yet the message from nature here is clear – there is a need in the design of new materials to develop mechanisms at a multiple of length scales in order to create new hybrid materials with unique mechanical properties.

Nacre is the iridescent material that thickly coats many shells and molluscs, such as the abalone shell, and which has long attracted the attention of materials scientists. The structural part of interest, on the basis of its astonishing properties, is the layer of nacreous aragonite covering the inside of the shell. Nacre is a nanocomposite composed of 95% calcium carbonate, one of the most abundant – but also one of the weakest – minerals on Earth. In spite of this intrinsic weakness, the toughness of nacre is 1000 to 3000 greater than that of the bulk monocrystalline calcium carbonate [14]; such dramatic improvement is obtained through a complete integration of the components in the structure of nacre.

Nacre exhibits a complex hierarchical architecture spanning several length scales. At the micrometer scale, nacre appears as a stunningly arranged layered material (Fig. 1), with the layers being composed of individual mineral platelets each of approximately 5 µm in diameter and 0.5 µm in thickness. Nano-asperities are covering the surface of the platelets, providing a specific surface roughness, and a few nanoscale-sized columns are found in the organic matrix layers. The aragonite columns, which in biomineralization are traditionally referred to as mineral bridges, pass through the mortar layers from one platelet to another and appear to be almost circular. Interlocks are also observed between the platelets of nacre; when two platelets are stacked one above another, the upper platelet with organic material penetrates into the lower platelet. The depth of the interlocking feature was found, on average, to be 20% of the thickness of the platelets. The tablets of nacre are constructed from a continuous organic matrix, which breaks the mineral up into coherent nanograins (20- to 50-nm size) which share the same crystallographic orientation, providing increased cohesion between the grains. The mineral platelets are separated by thin, 10- to 30-nm-thick layers of biological organic adhesive composed of polysaccharide and protein fibres (Fig. 1); this represents the remaining 5 vol% of the material, and contributes extensively to the unique physical and mechanical properties of nacre.

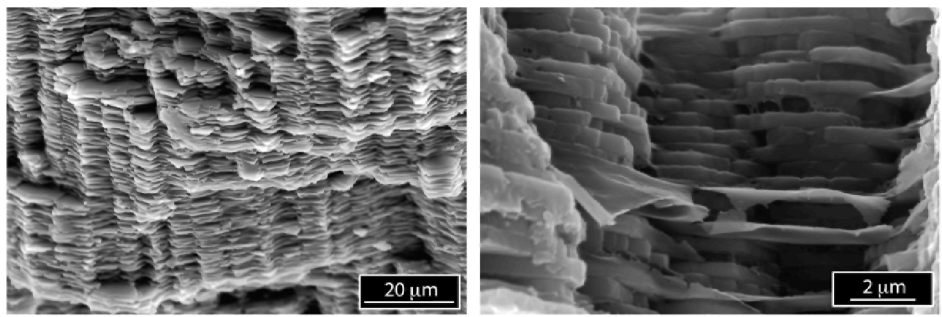

**FIGURE 1**. Structure of nacre, showing the multilayered structure (left) and the thin organic layer of proteins found between the calcium carbonate platelets (right).

The unique properties of natural layered materials and nanocomposites are achieved through a fine control of the layers thickness, selection of the correct components, and manipulation of the roughness and adhesion at the organic–inorganic interface. Although built in soft conditions (ambient temperature) and from a narrow selection of intrinsically weak materials (phosphates, carbonates), the functional properties exhibited by biological composites often surpass those of synthetic materials. The example of nacre is far from being unique, and a large body of such examples has been identified, with structures optimized for a variety of functional properties such as strength and stiffness (nacre), optical properties (Euplectella sponge) or adhesion (gecko).

Biological structures have hence long attracted the attention of scientists and engineers as fine blueprints to guide the design of new advanced materials and overcome current limitations of synthetic materials.

## Ice as a Natural Template

In sea ice, pure hexagonal ice platelets with randomly oriented horizontal $c$ crystallographic axes are formed, and the various impurities originally present in sea water (salt, biological organisms, etc.) are expelled from the forming ice and entrapped within channels between the ice crystals [15]. We applied this natural principle to ceramic particles dispersed in water to build sophisticated, nacre-like architectures in a simple two-step approach.

The physics of water freezing has drawn the attention of scientists from a wide range of specialties for a long time. With few exceptions, much of this work has concentrated on the freezing of pure water or very dilute suspensions [16-18]. This phenomenon is critical for various applications, such as cryo-preservation of biological cell suspensions and the purification of pollutants. A common conclusion of these studies is the presence of a critical particle size during the freezing of such suspensions, size above which the suspended particles will be trapped by the moving water-ice front. For a size below the critical size, particles are rejected by the solidification front and entrapped between the growing ice crystals. Another important observation is that the hexagonal ice crystals exhibit strong anisotropic growth kinetics, varying over about two orders of magnitude with crystallographic orientation; growth occurs very fast in the basal plane. Under steady-state conditions, it is thus possible to grow ice crystals in the form of platelets, with a very high aspect ratio; the c-axes of ice corresponds to the direction of the small thickness of the platelet. The ice thus formed will have a lamellar microstructure, with the lamellae thickness depending mainly on the speed of the freezing front. If these crystals are grown in a colloidal suspension of ceramics (such as alumina), the colloidal particles will end up entrapped in between the lamellar ice, forming themselves lamellar bodies of packed particles.

Ice is then removed by sublimation at low temperature and pressure, avoiding the drying stresses and shrinkage that may lead to cracks and warping during normal drying, and yielding porous green bodies that can be sintered in a final step. Scaffolds thus obtained can be used as-is, with a complex porous structure being the replica of the ice scaffolds before sublimation, or be filled with a suitable secondary phase to obtain dense composites with a complex multilayered structure.

We show here how this freezing technique can be applied not only to alumina (a model ceramic material) but also to hydroxyapatite, an osteophilic ceramic related to the inorganic component of bone, to process bone substitute materials with suitable physical and mechanical properties. In particular, we describe here how the processing conditions (concentration, freezing rate, sintering) affects the scaffold characteristics (size and amount of porosity, compressive strength) and discuss the potential and the limits of the technique.

## MATERIALS AND METHODS

Slurries were prepared by mixing distilled water with a small amount (1 wt% of the powder) of ammonium polymethacrylate anionic dispersant (Darvan C or 811, R. T. Vanderbilt Co., Norwalk, CT), an organic binder (polyvinyl alcohol, 2 wt.% of the powder) and the hydroxyapatite (Hydroxyapatite#30, Trans-Tech, Adamstown, MD) or alumina powder (Ceralox SPA05, Ceralox Div., Condea Vista Co., Tucson, USA) in various proportions. Slurries were ball-milled for 20 hrs with alumina balls and de-aired by stirring in a vacuum desiccator, until complete removal of air bubbles (typically 30 min). The alumina powder used in the study has a specific area of 8.1 $m^2/g$ and an average grain size of 400 nm (data provided by manufacturer). The hydroxyapatite powder used in the study has an average grain size of 2 microns (data provided by manufacturer).

Freezing of the slurries was done by pouring them into a teflon mold (18 mm diameter, 30 mm length) with two copper rods on each side which were cooled using liquid nitrogen. Freezing kinetics were controlled by heaters placed on the metallic rods and thermocouples placed on each side of the mold. Freezing occurred from bottom to top of the sample. Frozen samples were freeze-dried (Freeze dryer 8,

Labconco, Kansas City, MI) for 24 hours. By adjusting both the temperature gradient and the cooling rate, a wide range of freezing conditions can be investigated. For low freezing rates, only the bottom cold finger was used. To reach higher freezing rates, a constant macroscopic temperature gradient was established using the two cold fingers cooled at the same rate. The average ice front velocity was estimated by measuring the time of freezing and dividing by the length of the frozen sample.

The green bodies thus produced were sintered in air for 2 hours at 1500°C (alumina) or in the temperature range 1250-1350°C (hydroxyapatite), with heating and cooling rates of 5°C/min (1216BL, CM Furnaces Inc., Bloomfield, NJ). The microstructure of the samples was analyzed by optical and environmental scanning electron microscopy (ESEM, S-4300SE/N, Hitachi, Pleasanton, CA) and their total porosity was derived from the apparent density, measured by Archimedes's method. The wavelength was measured at the top 2/3 of the sample where a constant thickness of the ceramic lamellae was reached. The wavelength was measured on lines perpendicular to the ceramic lamellae. More than 100 measurements per sample were performed.

## RESULTS AND DISCUSSION

### Microstructure

As the particles are being rejected and trapped between the growing ice crystals during freezing of the slurries, the porosity in the sintered materials is a direct replica of the ice crystal before sublimation. This should be kept in mind while describing the specific structure of these materials and provide abundant information on the pattern formation mechanisms and means of control over the structure.

#### *A Layered Material*

The structure can be globally described as porous and multilayered, to some extent very similar to that obtained by freeze casting of polymeric materials, e.g., for tissue engineering. The structure is built from ceramic (alumina or hydroxyapatite) plates with flat interconnected macropores between them, aligned along the ice growth direction (Fig. 2). Colonies of locally aligned pores are observed, and some long-range order is usually found. Channels run continuously from the bottom to the top of the sample (samples were frozen from the bottom to the top). On the internal walls of the plates (or lamellae), a dendritic, branch-like structure can be observed following the microscopic ice formation. The dendritic surface relief covers only one side of the ceramic plates. The ceramic plates separating the lamellar channels are completely dense after sintering, with almost no visible residual porosity. The sintered scaffolds had total porosities between 30 and 70% depending on the solids concentration in the initial slurries and the width of the open interconnected macropores, which typically ranges between 20 and 100 μm in their smallest dimension and 50 to 500 μm in the largest one.

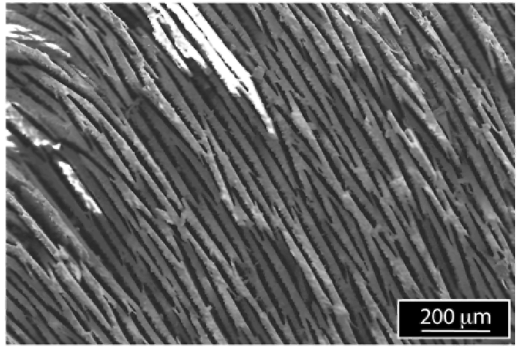

**FIGURE 2.** Lamellar structure of porous alumina.

## Nacre-like Composites

The inorganic portion represents 95% of the volume of nacre, but its highly specific properties (in particular its large toughness) are due to the interaction of this inorganic component with the organic phase (proteins) found between the calcium carbonate platelets. To obtain similar synthetic materials, the porous scaffolds can be filled with a second phase. For example, we filled the ice templated scaffolds with a simple organic phase (epoxy) or with an inorganic component (aluminum). Such materials exhibit strong similarities with the structure of natural materials such as nacre, bone or tooth (Fig. 3). We choose epoxy and aluminum as proofs of principles materials, but any suitable materials can be infiltrated, providing the functional requirements (surface energy, viscosity) are fulfilled. In particular, biopolymers (PLA, PCL, collagen) can be considered as secondary phase for biomaterials applications.

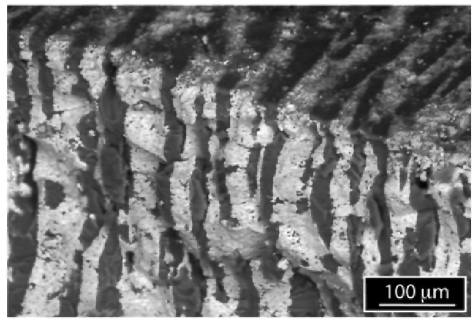

**FIGURE 3.** Example of hydroxyapatite/epoxy composite.

## A Hierarchical Material

Looking carefully at the structure of the processed samples, it appears to be more than just a lamellar structure. Several characteristic structural features were observed, each at a different length scale. The high level of hierarchical organization goes from the meso- (millimeters) down to the atomic level. The following organization has been observed:
- Macrostructure: the orientation of the lamellae can exhibit some long range preferential orientation. It can be made parallel, circular, or any other gradient of orientation can be considered.
- Lamellar structure, as described previously. Alternative morphologies can also be observed under certain conditions.
- Dendritic structure covering the surface of the ceramic lamellae, as described previously (Fig. 4).

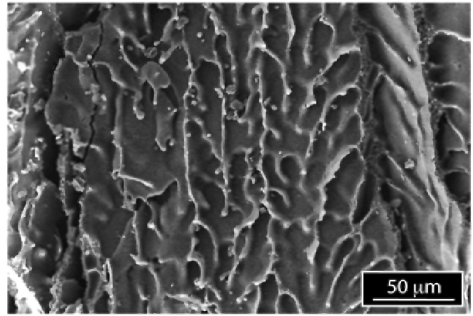

**FIGURE 4**. Dendritic roughness covering the ceramic lamellae (alumina).

- Ceramic bridges linking adjacent lamellae. In the sintered porous structures, these numerous fine features with often-tortuous morphologies are locally bridging the gap between the two adjacent lamellae. The morphology of these features is very different from that of the dendrites covering the ceramics lamellae, suggesting another formation mechanism (Fig. 5).

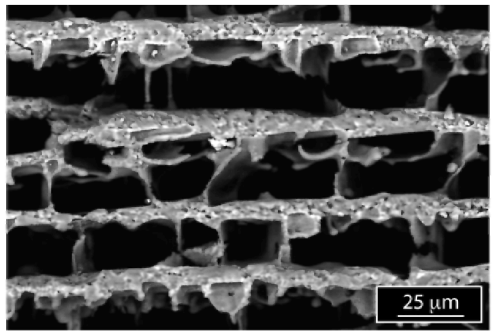

**FIGURE 5**. Ceramic bridges linking adjacent ceramic lamellae (alumina).

- Submicronic grains, composing the lamellae and the bridges (Fig. 6). The size of the grains is directly related to the sintering conditions.

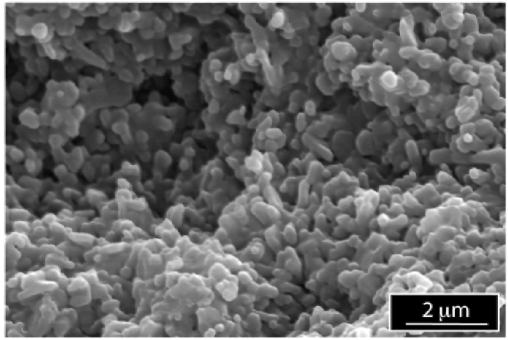

**FIGURE 6.** Submicronic alumina grains composing the ceramic lamellae.

- Nanoscopic reinforcements. These reinforcements can be added to the structure at several stages. One easy way to achieve this is to graft the initial alumina powder with zirconia precursors (zirconium alkoxyde), such that a zirconia-toughened alumina nanocomposites are obtained (Fig. 7).

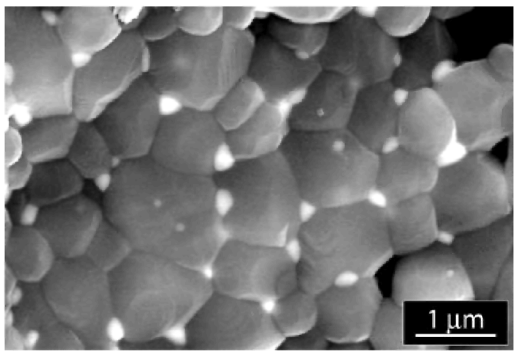

**FIGURE 7**. Nanoscopic reinforcements: zirconia grains at triple point in the alumina matrix composing the ceramic lamellae.

- Grain boundaries. They might be controlled (by dopants, additives) to tailor some functional properties like creep or conductivity.

In addition, two other degrees of hierarchy can be considered when composites are made from the ice-templated porous scaffolds, i.e., after infiltration with a suitable second phase (polymer or metal).
- Nature of the infiltrated phase. The nature and properties of the infiltrated phase will be critical in regards of the properties exhibited by such composites.
- Interface between the ceramic phase and the secondary phase. The issue of interfaces in these materials can be particularly significant and is illustrated and discussed later on in the paper.

# Properties

## *Porous Materials*

Compressive strength vs. total porosity for the hydroxyapatite scaffolds revealed some dramatic improvements, in comparison to existing materials solutions [19,20]. Although the strength for high-porosity content (typically >60 vol%) is comparable to that reported in the literature, it increases rapidly when the porosity decreases. Values obtained for these samples are well above those reported so far in the literature. It has previously been shown how the cooling rate strongly affects the porosity features (size and morphology). These changes in cooling rate have, consequently, a direct repercussion on the compressive strength. For low cooling rate (<2°C/min), the compressive strength remains moderate, at 10 to 20 MPa. However, for fast cooling rates (>5°C/min), the compressive strength increases up to 60 MPa as the cooling rate increases.

The presence of inorganic bridges between the ceramic lamellae (a feature that parallels the microstructure of nacre) prevents Euler buckling of the ceramic lamellae and contributes to the high strength. In fact, the strength of the porous lamellar HAP is similar to that of compact bone. Load-bearing biological applications, requiring high strength, might now be considered with such materials.

## *Dense Composite Materials*

The composites obtained after epoxy or aluminum infiltrations were tested in three-point bending to investigate the crack propagation mechanisms. Nature shows that the optimum fracture properties are encountered not only when the organic/inorganic interface is strong, but also when delamination at the organic/inorganic interface occurs before the crack goes across the stiff, brittle layer. Extensive crack deflection at the interface between lamellae was observed (Fig. 8) in the ice templated composites. As in nacre, this delamination creates tortuous cracks, which propagate in a stable manner and increase the toughness of the materials. Multiple cracking, and bridging of the main crack, were also observed in the metal/ceramic composites.

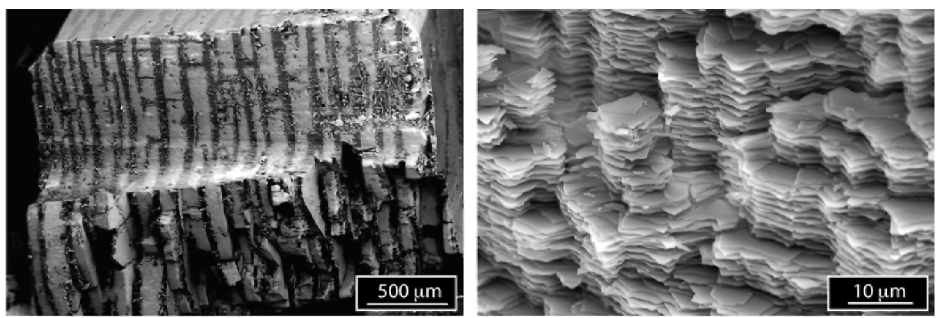

**FIGURE 8**. Fracture surface of a nacre-like alumina-epoxy sample (left). A clear delamination along the interface is observed, similar to what is observed in nacre (right), and yielding improved crack propagation resistance and toughness properties.

The load-displacement curves of the tested samples [19] revealed a gradually decreasing load, very similar to that observed in the testing of nacre. This behavior can be directly related to a stable crack propagation perpendicular to the layer with active toughening. Even if the toughness value of the HAP composite (220 J/m$^2$) is still much lower than that of nacre (1000-3000 J/m$^2$) —partly because epoxy has a low toughness in comparison with the protein layer found in nacre— the crack propagation mechanisms seem very similar. This type of microstructure can therefore offer promising perspectives for increasing toughness in ceramic composites. A deeper characterization of the mechanical properties of these materials is currently being performed.

## Crystal Growth and Patterns Formation: Control of the Structure

The fact that a lamellar structure is obtained is intrinsically related to the nature of the solvent, ice, and its physics. The pattern formation in the ceramic body is directly related to the pattern formation of ice crystals. Several examples are described below.

The first two examples are related to the underlying mechanisms controlling the formation of the porous structure. In order to obtain ceramic samples with a lamellar porous structure, two requirements must be satisfied:

(1) The ceramic particles in suspension in the slurry must be rejected from the advancing solidification front and entrapped between the growing ice crystals.
(2) The ice front must have a columnar or lamellar morphology.

- *Interaction with the particles*: The physics of water freezing has drawn the attention of scientists for a long time. With few exceptions, much of this work has concentrated on the freezing of pure water or very dilute suspensions. This phenomenon is critical for various applications, such as cryo-preservation of biological cell suspensions and the purification of pollutants. An important observation in these studies is that, during the freezing of such suspensions, there is a critical particle size for a given velocity (or critical velocity for a given particle size) above which the suspended particles will be trapped by the moving water-ice front. Typical critical velocities vary from 1 to 10 μm/s for particles with a diameter ranging from 1 to 10 μm. Since the critical velocity is inversely proportional to the particle radius, it is likely to be very high (>100 μm/s) for submicronic particles such as the ones used here. Hence, under the usual slurries formulations and freezing conditions, the ceramic particles are always rejected by the moving ice front and entrapped between the ice crystals.

- *Lamellar morphology*: The ice formed under these conditions of pressure and temperature has a hexagonal crystallographic structure, exhibiting a strong anisotropy of growth kinetics. The ice front velocity parallel to the crystallographic c-axis is 10$^2$ to 10$^3$ times lower than perpendicular to this axis (in the basal plan). Ice platelets with a very large anisotropy can then be formed very fast with ice growing along the *a*-axes, while the thickness (along the *c*-axis) remains small. The freezing process is easier for crystals whose *c*-axes are perpendicular to the temperature gradient, such that growth along the gradient can occur in the *a*- or *b*-direction. The crystals with horizontal *c*-axes will therefore grow at the expense of the others and continue to grow upward, in an architecture composed of long vertical lamellar crystals with horizontal *c*-axes. In the final structures, the direction perpendicular to the lamellae corresponds thus to the original *c*-axis of ice crystals.

Besides, these properties can also be used to explain the observed relationships between the freezing kinetics and the ceramic lamellae thickness [19,21]. When the freezing kinetics is increased, i.e., the solidification front speed increases, which in turn reduces the thickness of the lamellar ice crystals, and hence the thickness of the ceramic lamellae in the final material. By modifying the freezing kinetics, the thickness has been adjusted in the range 2 to 200 microns [19]. Further progresses can be expected with improved experimental setups. For a deeper discussion on this dependency, the reader can refer to reference [21].

- *Initial Gradient*: Some time is necessary for the freezing front to evolve from a planar to a columnar or lamellar/dendritic morphology. The transition is reflected in the final architecture of the porous structures, in the first frozen zone, initially close to the cold finger (Fig. 9) [20]. The first frozen zone reveals a planar ice front where the alumina particles were entrapped. The interface then moves progressively to a columnar and eventually lamellar morphology, with a progressive ordering of the lamellae. A steady state is eventually reached and ice crystals become continuous, running through the entire sample, with a constant thickness, yielding a porous structure with a morphology that has been described previously.

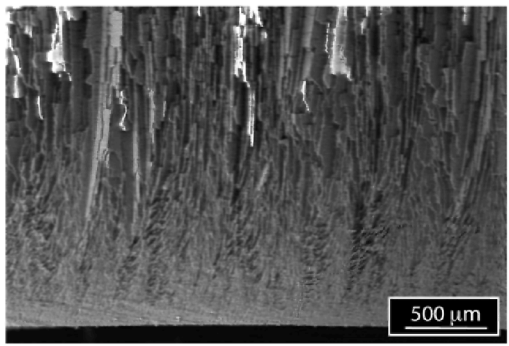

**FIGURE 9.** Initial gradient of structure in the sample (alumina) with a gradual change in the morphology of the porosity, corresponding to the initial unstable solidification regime. Freezing occurred from the bottom to the top.

# CONCLUSIONS AND PERSPECTIVES

The promise of layered materials has not been fully achieved because we do not have the technology to fabricate the needed layers in bulk materials. Although the message from biology and natural materials is clear –there is a need to control the structure of the material at several length scales to provide the desired functional properties– suitable processing techniques are still lacking.

We summarized here a very simple and natural technique allowing us to easily process bulk layered materials with a hierarchical structure relevant length scales, and control their properties into ways that parallel the structure of nacre. Dramatic improvements of functional properties were measured on such materials.

The pattern formation in the ceramic body is directly related to the pattern formation of ice crystals, and the knowledge of the underlying physical principles provides numerous powerful means of controls over the process and the resulting structures. In particular, the thickness of the layers can be varied over two orders of magnitude by adjusting the freezing kinetics.

The segregation phenomenon occurring during freezing is based on physical interactions (not chemical). Therefore, any kind of ceramic particles can be used, making the process a very versatile one. Our technique shows promise for a large number of applications that require tailored composite materials. One such scientific challenge that could be solved is the development of new biomaterials for orthopaedic applications. Despite extensive efforts in the development of porous scaffolds for bone regeneration, all porous materials have an inherent lack of strength associated with porosity. By applying freezing to commercial hydroxyapatite (HAP, the mineral component of bone) powder suspensions, we processed ice templated highly porous lamellar scaffolds that are four times stronger in compression than conventional porous HAP. These materials could be considered as potential candidates for bone and teeth applications.

## ACKNOWLEDGMENTS

This work was supported by the National Institute of Health (NIH/NIDCR) under grant No. 5R01 DE015633 (Complex Nanocomposites for Bone Regeneration) and by the Director, Office of Science, Office of Basic Energy Sciences, Division of Materials Sciences and Engineering of the Department of Energy under Contract No. DE-AC03-76SF00098 (Metal/Ceramic Composites).